\documentclass[12pt]{article}
\usepackage{amsmath,amsfonts,amssymb,dcolumn,lscape,subfigure,alltt,authblk}
\usepackage{graphicx,color}
\usepackage{tabularx}
\begin{document}

\baselineskip=5.5mm
\setlength\parindent{0pt}
\newcommand{\be} {\begin{equation}}
\newcommand{\ee} {\end{equation}}
\newcommand{\Be} {\begin{eqnarray}}
\newcommand{\Ee} {\end{eqnarray}}
\renewcommand{\thefootnote}{\fnsymbol{footnote}}
\def\a{\alpha}
\def\b{\beta}
\def\g{\gamma}
\def\G{\Gamma}
\def\d{\delta}
\def\D{\Delta}
\def\e{\epsilon}
\def\k{\kappa}
\def\l{\lambda}
\def\L{\Lambda}
\def\t{\tau}
\def\om{\omega}
\def\Om{\Omega}
\def\s{\sigma}
\def\lg{\langle}
\def\rg{\rangle}
\def\G{{\bf G}}
\author[]{Gregor Diezemann}
\affil[]{Institut f\"ur Physikalische Chemie, Universit\"at Mainz, Duesbergweg 10-14, 55128 Mainz, Germany}
\title
{Nonlinear response theory for Markov processes II: Fifth-order response functions}
\maketitle
\begin{abstract}
The nonlinear response of stochastic models obeying a master equation is calculated up to fifth-order in the external field thus extending the third-order results obtained earlier (G. Diezemann, Phys. Rev. E{\bf 85}, 051502 (2012)).
For sinusoidal fields the $5\om$-component of the susceptibility is computed for the model of dipole reorientations in an asymmetric double well potential and for a trap model with a Gaussian density of states.
For most realizations of the models a hump is found in the higher-order susceptibilities.
In particular, for the asymmetric double well potential model there are two characteristic temperature regimes showing the occurence of such a hump as compared to a single characteristic regime in case of the third-order response.
In case of the trap model the results strongly depend on the variable coupled to the field.
As for the third-order response, the low-frequency limit of the susceptibility plays a crucial role with respect to the occurence of a hump.
The findings are discussed in light of recent experimental results obtained for supercooled liquids.
The differences found for the third-order and the fifth-order response indicate that nonlinear response functions might serve as a powerful tool to discriminate among the large number of existing models for glassy relaxation.
\end{abstract}
\newpage
\section*{I. Introduction}
The nonlinear response of supercooled liquids and glasses to electrical fields has been investigated intensively during the last decade both, experimentally and theoretically
\cite{CrausteThibierge10,Brun11,Lunki17}.
Motivated by a relation between the maximum value of the modulus of the cubic susceptibility and the volume of correlated domains\cite{Bouchaud05} the data have been mostly analyzed accordingly and the number of correlated particles in a typical domain, $N_{\rm corr}$, have been extracted.
The values obtained are compatible with what is to be expected if one assumes a growing length scale to be responsible for the heterogeneous slow dynamics of glass-forming 
liquids\cite{SRS91,HWZS95,G23,Sillescu99,Ediger00,Richert02,Toninelli05,Berthier:2011}.

Experimentally, a number of different techniques have been applied to obtain dynamical nonlinear susceptibilities\cite{Lunki17}.
In particular, and most important for the present work, it has been found that the scaled modulus of the 
3$\om$-response $\chi_3^{(3)}(\om,T)$,
\be\label{X3.Def}
\hat X_3(\om,T)=\left|\chi_3^{(3)}(\om,T)\right|{k_BT\over(\D\chi_1)^2a^3}
\ee
where $\D\chi_1$ denotes the static linear response, $k_B$ the Boltzmann constant and $a^3$ the molecular volume, exhibits a hump-like structure.
It is the height of this hump that has been related to $N_{\rm corr}$ for supercooled 
liquids\cite{CrausteThibierge10,Bauer:2013,Casalini:2015}.
Such a hump has not only been observed in molecular glass forming liquids, but also in a particular alcohol\cite{Bauer:2013}, in plastic crystals\cite{Michl:2015,Michl:2016} and a ionic 
liquid\cite{Patro:2017}.

There have been some model calculations that aim at a quantitative description of the data obtained.
The so-called box model was found to account for some of the features 
observed\cite{Brun11b,Pick:2015}.
A modification of a similar model shows results consistent with the experimental ones without the assumption of any spatial correlations\cite{Richert:2016}.
A toy model considers a number of dipoles reorienting in an asymmetric double well potential and includes the so-called trivial reorientations at low frequencies\cite{Ladieu:2012}.

The nonlinear response theory of dipole reorientations has been worked out for various models already some time ago\cite{Morita86,DD95,Dejardin00,Kalmykov01} and has also been extended to include long-range dipolar interactions\cite{Dejardin:2014}.
In ref.\cite{G75}, denoted as I in the following, I have computed the third-order nonlinear response for Markov processes using time-dependent perturbation theory for the propagator of the master equation (ME) governing the stochastic dynamics.
In particular, I considered two simple stochastic models and computed the modulus of the cubic response for these.
One model considers dipole reorientations in an asymmetric double well potential, described as a two-state model 
(ADWP model)\cite{G39,G46}.
This model shows a hump in the reduced modulus in a narrow temperature range located around a characteristic temperature $T_3$ at which the low-frequency response vanishes.
For temperatures below $T_3$, the height of the hump increases with temperature at variance with experimental findings. 
In the range $T>T_3$, the height decreases with increasing $T$.
The other model I considered is the trap model with a Gaussian density of traps\cite{Dyre95,MB96,G64,G71,G73}.
Depending on the particular choice of the variable that couples to the field a hump is obtained in most cases the height of which shows different temperature dependences.
For specific variables it increases, for others it decreases or is essentially temperature-independent.
The conclusion from these calculations is that it is indeed possible to obtain a hump in the modulus of the third-order response from such mean-field models without assuming the existence of glassy correlations.

Quite recently, the fifth-order susceptibility ($\chi_5$) of glycerol and propylene carbonate has been measured and also in this case a hump at a frequency somewhat below the inverse $\a$-relaxation time has been observed. 
The relation between the relative moduli has been interpreted in terms of the existence of compact dynamically correlated 
regions\cite{Albert:2016}.
In addition, it is argued that a number of models for slow dynamics cannot account for the observed features, including the box model.
This is very interesting because, as noted above, the box model has been shown to be able to describe the third-order 
response\cite{Brun11b,Pick:2015} and it also has been used successfully to analyze nonlinear hole-burning data\cite{SBLC96,G16}.
Therefore, it appears possible that measurements of higher-order nonlinear response functions might provide additional information that allows to discriminate among different models for the slow relaxation in glass-forming liquids.

In the present paper, I will compute the fifth-order response function $\chi_5^{(5)}(\om)$ for the same models considered in I.
In particular, I will discuss the effect of sinusoidal fields of the form $H(t)=H_0\cos{(\om t)}$.
For this oscillating field the various response functions for times long compared to the initial transients can be written as:
\Be\label{Chi.om.def}
\chi^{(1)}(t)
&&\hspace{-0.6cm}=
{H_0\over2}\left[e^{-i\om t}\chi_1(\om)+c.c.\right]
\nonumber\\
\chi^{(3)}(t)
&&\hspace{-0.6cm}=
{H_0^3\over2}\left[e^{-i\om t}\chi_3^{(1)}(\om)+e^{-i3\om t}\chi_3^{(3)}(\om)+c.c.\right]
\\
\chi^{(5)}(t)
&&\hspace{-0.6cm}=
{H_0^5\over2}\left[e^{-i\om t}\chi_5^{(1)}(\om)+e^{-i3\om t}\chi_5^{(3)}(\om)+e^{-i5\om t}\chi_5^{(5)}(\om)+c.c.\right]
\nonumber
\Ee
where $c.c.$ denotes the complex conjugate.

In the following sections, I will mainly discuss the scaled moduli of the maximum frequency-component,
\be\label{X35.Def}
X_3(\om,T)={T\over(\D\chi_1)^2}\left|\chi_3^{(3)}(\om,T)\right|
\quad\mbox{and}\quad
X_5(\om,T)={T^2\over(\D\chi_1)^3}\left|\chi_5^{(5)}(\om,T)\right|
\ee
The prefactors $T/(\D\chi_1)^2$ and $T^2/(\D\chi_1)^3$ eliminate the trivial temperature dependences stemming from 
$\D\chi_1\sim T^{-1}$, and $\chi_3^{(3)}\sim T^{-3}$, $\chi_5^{(5)}\sim T^{-5}$.
This allows to compare the different moduli directly.
As outlined in the following section, the calculations are performed employing perturbation theory for the propagator of a master equation for Markov processes.
\section*{II. Fifth-order response functions}
Here, I briefly discuss the calculation of the response of a dynamical system with a kinetics described by a master equation\cite{vkamp81} using the same notation as in I\cite{G75}.

Writing $G_{kl}(t,t_0)$ for the conditional probability to find the system in state $k$ at time $t$ provided it was in state $l$ at time $t_0$, the ME has the form
\be\label{ME.t.abh}
\dot G_{kl}(t,t_0)=-\sum_nW_{nk}(t)G_{kl}(t,t_0)+\sum_nW_{kn}(t)G_{nl}(t,t_0)
\ee
where the rates for a transition from state $k$ to state $l$ are given by $W_{lk}(t)$.
The response of the system to an external field applied at time $t_0$ and measured by an observable $F(t)$, 
\be\label{F.expect}
\lg F(t)\rg_{(H)}=\sum_{kl}F_kG^{(H)}_{kl}(t,t_0)p_k(t_0)
\ee
is determined by the one-time probabilities $p_k(t)$ obeying the same ME and given by $p_k(t)=\sum_lG_{kl}(t,t_0)p_l(t_0)$.
Here, $G^{(H)}_{kl}(t,t_0)$ denotes the propagator in the presence of the field and for the transition rates I use 
\be\label{Wkl.HX}
W_{kl}^{(H)}(t)=W_{kl}(t)e^{\b H[\g M_k-\mu M_l]}
\ee
where $\g$ and $\mu$ can be chosen arbitrarily\cite{G39,CR03,G54} and $\beta=T^{-1}$ with the Boltzmann constant set to unity.
However, usually one considers models that obey detailed balance and therefore relates the parameters via $\g=1-\mu$.

A perturbation expansion is achieved via the expansion of the transition rates $W_{kl}^{(H)}(t)$ and a concomittant expansion of $\G^{(H)}(t,t_0)$ in terms of the corresponding 'field-free' propagator $\G(t,t_0)$ using the decomposition
${\cal W}^{(H)}(t)={\cal W}(t)+{\cal V}(t)$ with ${\cal V}(t)=\sum_{n=1}^\infty{\cal V}^{(n)}(t)$, cf. I.

The perturbation expansion for the propagator starts from the Dyson-like equation
\be\label{G.Dyson}
\G^{(H)}=\G+\G\otimes{\cal V}\otimes\G^{(H)}
\ee
where I omitted the time arguments and the convolution is abbreviated by
$\G\otimes{\cal V}\otimes\G^{(H)}\equiv\int_{t_0}^t\!dt'\G(t,t'){\cal V}(t')\G^{(H)}(t',t_0)$.

Using the series expansion for the transition rates, one easily finds for the relevant terms:
\Be\label{G.bis5}
\G^{(1)}
&&\hspace{-0.6cm}=
\G\otimes{\cal V}^{(1)}\otimes\G
\nonumber\\
\G^{(3)}
&&\hspace{-0.6cm}=
\G\otimes{\cal V}^{(3)}\otimes\G
+\left\{\G\otimes{\cal V}^{(1)}\otimes\G\otimes{\cal V}^{(2)}\otimes\G\right\}
\nonumber\\
\G^{(5)}
&&\hspace{-0.6cm}=
\G\otimes{\cal V}^{(5)}\otimes\G
  +\left\{\G\otimes{\cal V}^{(1)}\otimes\G\otimes{\cal V}^{(4)}\otimes\G\right\}
\nonumber\\
&&\hspace{2.4cm}
+\,\left\{\G\otimes{\cal V}^{(2)}\otimes\G\otimes{\cal V}^{(3)}\otimes\G\right\}
\\
&&\hspace{2.4cm}
+\,\left\{\G\otimes{\cal V}^{(1)}\otimes\G\otimes{\cal V}^{(1)}\otimes\G\otimes{\cal V}^{(3)}\otimes\G\right\}
\nonumber\\
&&\hspace{2.4cm}
+\,\left\{\G\otimes{\cal V}^{(1)}\otimes\G\otimes{\cal V}^{(2)}\otimes\G\otimes{\cal V}^{(2)}\otimes\G\right\}
\nonumber\\
&&\hspace{2.4cm}
+\,\left\{\G\otimes{\cal V}^{(1)}\G\otimes{\cal V}^{(1)}\otimes\G\otimes{\cal V}^{(1)}\otimes\G\otimes{\cal V}^{(2)}\otimes\G\right\}
\nonumber\\
&&\hspace{2.4cm}
+\,\,\G\otimes{\cal V}^{(1)}\G\otimes{\cal V}^{(1)}\G\otimes{\cal V}^{(1)}\otimes\G\otimes{\cal V}^{(1)}\otimes\G
\otimes{\cal V}^{(1)}\otimes\G
\nonumber
\Ee
Here, $\left\{\G\otimes{\cal V}^{(x)}\otimes\G\otimes{\cal V}^{(y)}\otimes\G\otimes{\cal V}^{(z)}\otimes\G\right\}$ is a shorthand notation for the sum of all permutations of $x$, $y$ and $z$.

In the next step, one uses the expression for the matrix elements of $\G^{(n)}(t,t_0)$, $G^{(n)}_{kl}(t,t_0)$, in 
eq.(\ref{F.expect}) in order to compute the nth-order response.
For the calculation of the fifth-order response one then proceeds in the same way as in I\cite{G75}.
I will not dwell on the general results further, but discuss the response functions obtained for specific models in the following. 
\section*{III. The ADWP-model}
As in refs.\cite{G75,G46}, two dipole orientations characterized by polar angles $\theta_1=\theta$ and $\theta_2=\theta+\pi$ are assumed 
to describe the minima of the double well potential of the system.
The flips of the dipoles take place with rates $W_{12}=We^{-\b\D/2}$ and $W_{21}=We^{+\b\D/2}$, where $\D$ denotes the asymmetry and the bare  rate for $\D=0$ is given by $W$.
If no field is applied, the solution of the ME reads as $G_{kl}(t)=p_k^{\rm eq}\left(1-e^{-t/\t}\right)+\d_{kl}e^{-t/\t}$
with the relaxation rate $\t^{-1}=2W\cosh{\!(\b\D/2)}$ and the equilibrium populations $p_k^{\rm eq}=\t\cdot W_{kl}$.
The moment coupling to the field is $M_k=M\cos{\!(\theta_k)}$, i.e. $M_1=M\cos{\!(\theta)}$ and $M_2=-M\cos{\!(\theta)}$ where $M$ denotes the static molecular dipole moment.
Assuming isotropic distributions of the moments leads to an average over the angle $\theta$ and one has 
$\lg\cos^n{\!(\theta_k)}\rg=(n+1)^{-1}$ for $n$ even and $\lg\cos^n{\!(\theta_k)}\rg=0$ otherwise.

The linear response for the model is given by $\chi_1(\om)=\D\chi_1[1-i\om\t]^{-1}$ with
$\D\chi_1=\b(M^2/3)\left(1-\d^2\right)$ and $\d=\tanh{\!(\b\D/2)}$ and its properties have been discussed in I\cite{G75}.

Also the results for the third-order response have been presented there already and I repeat them in Appendix A for completeness.
One important finding is that the scaled modulus, eq.({\ref{X35.Def})
\[
X_3(\om,T)={\D\chi_3\over\b(\D\chi_1)^2}|S_3^{(3)}(\om\t)|
\]
with $\D\chi_3=\b^3(M^4/20)(1-\d^2)$ shows a hump in a finite temperature regime (cf. Fig.3 of I), determined by the vanishing of the static nonlinear susceptibility $\chi_3^{(3)}(\om\to0,T)$, eq.(\ref{Chi3.0.ADWP})
This determines the characteristic temperature 
\be\label{T3.def}
T_3=\D/\ln{[(\sqrt{3}+1)/(\sqrt{3}-1)]}\simeq\D/1.317.
\ee
The consequences of the existence of $T_3$ and the resulting hump in $X_3$ has been discussed in detail in I.

Another third-order response that has been investigated experimentally and that shows a very similar behavior as $X_3(\om)$ is the response quadratic to a dc field and linearly to an ac field, $\chi_{2;1}(\om)$\cite{LHote:2014}.
Without going into details regarding the behavior of this function here, I just mention that it can be written in the form
$\chi_{2;1}(\om)=\D\chi_3\cdot S_{2;1}(\om\t)$ and the scaled modulus is given by
$X_{2;1}(\om)=(T\D\chi_3)/(\D\chi_1)^2\left|S_{2;1}(\om\t)\right|$ with the spectral function given in eq.(\ref{S33.ADWP}).
Due to the low-frequency limit of $S_{2;1}(\om)$, $S_{2;1}(0)=(3\d^2-1)$, one finds that $X_{2;1}(0)$ vanishes at the same characteristic temperature as $X_3$, $T_{2;1}=T_3$.
Of course, this behavior is expected, as it is solely determined by the zero frequency limit of the respective spectral functions.

For the fifth-order response, one finds the following expression:
\be\label{chi5.ADWP}
\chi_5^{(5)}(\om)=\D\chi_5\times S_5(\om\t)
\quad\mbox{with}\quad
\D\chi_5=\left({M^6\over112}\right)\b^5(1-\d^2)
\ee
where I again have performed the isotropic average.
The spectral function $S_5(x)\equiv S_5^{(5)}(x)$ is given in Appendix A, eq.(\ref{S5.ADWP}).
In Fig.\ref{Fig1}, the fifth-order response $\chi_5^{(5)}(\om)$ is shown for different values of the asymmetry $\D$.
\begin{figure}[h!]
\centering
\includegraphics[width=8.0cm]{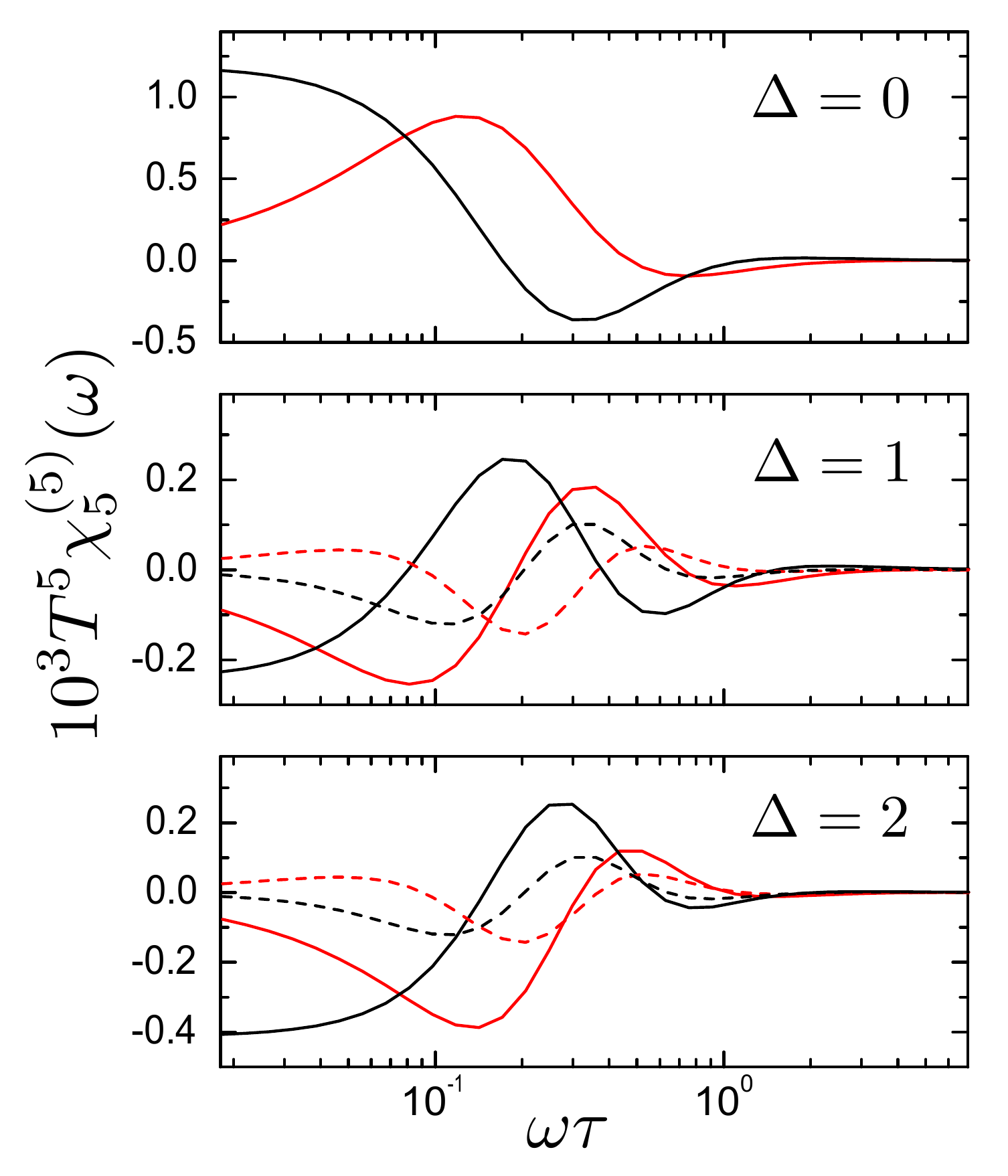}
\vspace{-0.5cm}
\caption{$\chi_5^{(5)}(\om)$ for various values of the asymmetry as indicated. The real part is represented by black lines and the imaginary part by red lines.
The full lines correspond to $T=1$ and the dashed lines to $T=T_{5;a}$.}
\label{Fig1}
\end{figure}
This can be compared directly to $\chi_3^{(3)}(\om)$ shown in Fig.2 of I.
The overall behavior is very similar.
The corresponding scaled modulus is given by, cf. eq.(\ref{X35.Def}):
\be\label{X5.ADWP}
X_5(\om,T)={|\chi_5^{(5)}(\om,T)|\over\b^2(\D\chi_1)^3}={\D\chi_5\over\b^2(\D\chi_1)^3}\left|S_5(\om\t)\right|
\ee
As in case of the third-order response functions, the modulus $X_5(\om,T)$ shows a hump in the temperature regime where the low-frequency limit of the susceptibility vanishes, i.e. where $S_5(0)=0$.
As this quantity is given by 
\[
S_5(0)=(2-15\!\cdot\!\d^2+15\!\cdot\!\d^4)/15
\]
one finds two characteristic temperatures:
\be\label{T5ab.def}
T_{5;a/b}=\D/\ln{[(1+z_{a/b})/(1-z_{a/b})]}
\quad\mbox{with}\quad
z_{a/b}=\sqrt{{15\mp\sqrt{105}\over30}}
\ee
which yields ($z_a\simeq0.398$, $z_b\simeq0.917$)
\be\label{T5ab.sim}
T_{5;a}\simeq\D/3.145
\quad\mbox{with}\quad
T_{5;b}\simeq\D/0.842
\ee
In Fig.\ref{Fig2} $X_5(\om)$ is shown in the two relevant temperature regimes centered around $T_{5;a/b}$ on a logarithmic scale.
As in case of the third-order response, a hump is observed in a rather narrow temperature range around $T_{5;a/b}$.
\begin{figure}[h!]
\centering
\includegraphics[width=8.0cm]{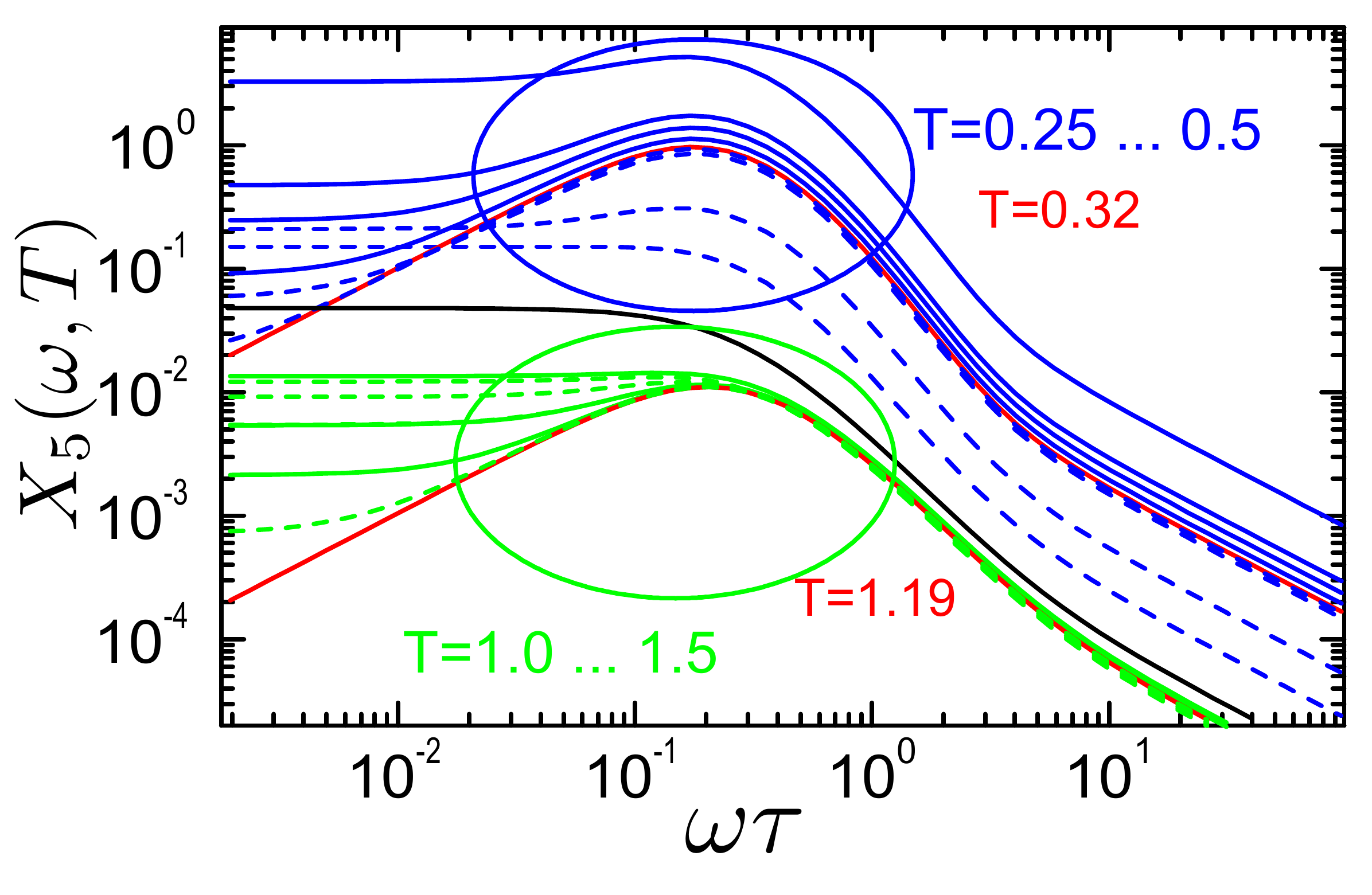}
\vspace{-0.5cm}
\caption{$X_5(\om)$ for $\D=1$ and various temperatures. 
The red lines correspond to $T=T_{5;a/b}$.
The black line corresponds to $T=T_3$.
The full lines represent temperatures higher than $T_{5;a/b}$ and the dashed lines are lower temperatures.
}
\label{Fig2}
\end{figure}
Note that in the temperature range around $T_3$ only trivial behavior is observed in $X_5$, cf. the black line in Fig.\ref{Fig2}.
This means that the experimental finding that both $X_3$ and $X_5$ show a hump in the investigated temperature regime with decreasing hump height with increasing temperature\cite{Albert:2016} cannot be reproduced by the ADWP model in its form used here.
This fact is further exemplified in Fig.\ref{Fig3}, where the relative height of the hump,
$X_k(\om_{\rm max})/X_k(0)$ for $k=3$, $5$ is shown.
\begin{figure}[h!]
\centering
\includegraphics[width=8.0cm]{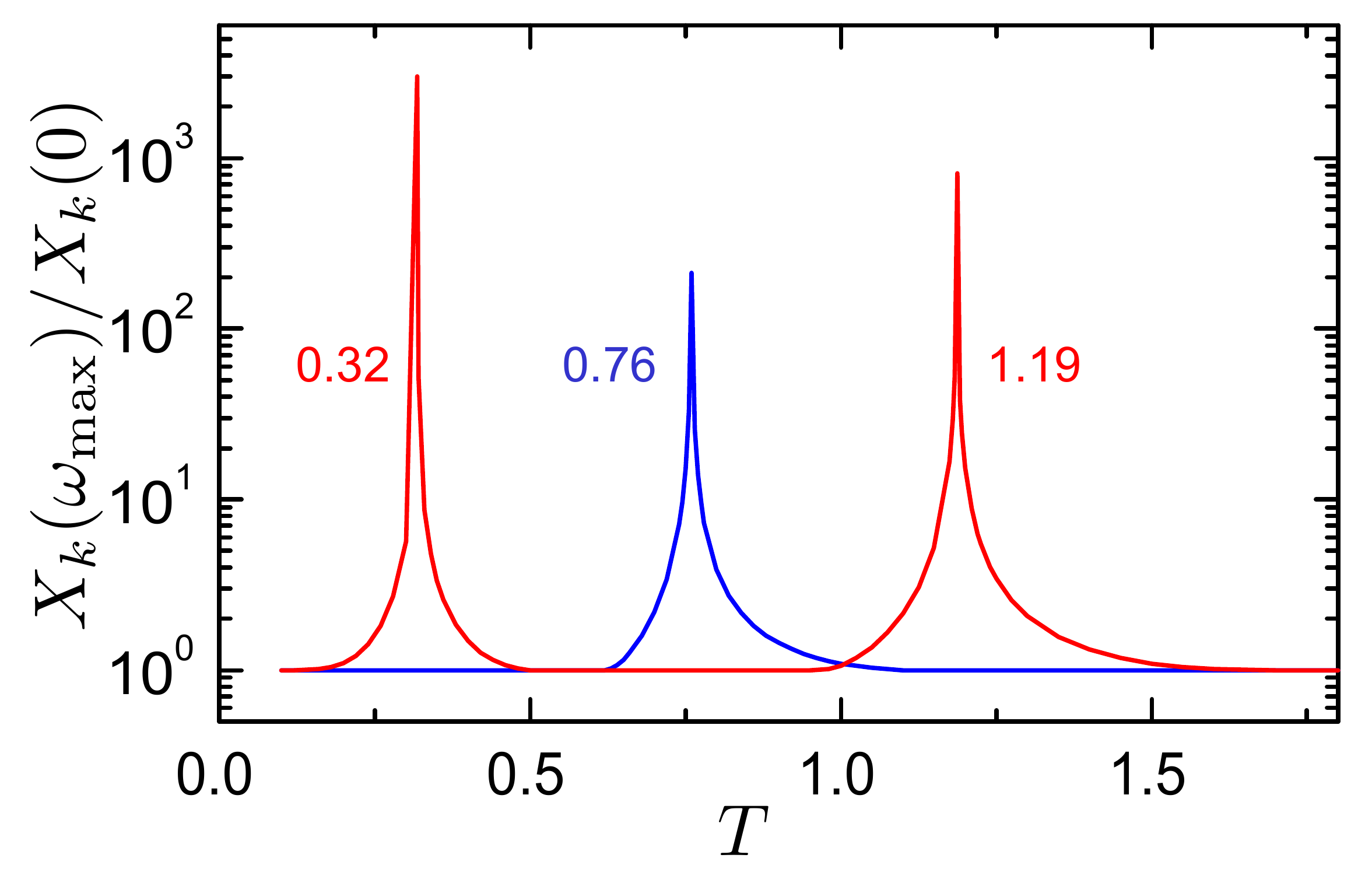}
\vspace{-0.5cm}
\caption{Relative height of the humps for the third-order (blue) and the fifth-order response (red) using 
$\D=1$.
}
\label{Fig3}
\end{figure}
One might argue that the experimental results for the third-order susceptibilities can be interpreted in terms of the ADWP model if one assumes that the relevant temperature range is above $T_3$.
On the basis of the calculated fifth-order response, one would additionally have to assume that the relevant temperature range is above $T_{5;b}$.
\section*{IV. The Gaussian trap model}
As in I, another model for glassy relaxation that I will consider is the well known trap model with a Gaussian density of 
states\cite{Dyre95,MB96,G64}. 
In brief, one considers the metastable states of a glass-forming liquid to be characterized by a free energy $\e$ and assumes transitions among different values of $\e$ to be given by 
\be\label{W.trap}
W(\e'|\e)=\rho(\e')\k(\e)
\quad\mbox{with the escape rate}\quad
\k(\e)=\k_\infty e^{\b\e}
\ee
This means that after an activated escape the destination trap is chosen at random, i.e. according to the density of states 
$\rho(\e')$.
The propagator $G(\e,t|\e')$ is obtained from the solution of the ME and all properties of the system can be obtained from this.
As in I, it is assumed that the field couples to a variable $M(\e)$, for which a Gaussian factorization is assumed to hold\cite{FS02}:
\be\label{M.eps.gauss}
\lg M(\e)\rg=0
\quad;\quad
\lg M(\e)M(\e')\rg=\d(\e-\e')e^{-n\b\e}
\ee
Here, $n$ is a model parameter.
Of course, this choice is quite arbitrary and can be relaxed.
However, eq.(\ref{M.eps.gauss}) has the advantage that the scaled linear response becomes independent of $n$, i.e. on the particularly chosen variable, cf. the discussion in I. 
As for the four point correlations needed for $X_3$, a Gaussian factorization property is assumed to hold for the six-point functions
$\lg M(\e_1)M(\e_2)M(\e_3)M(\e_4)M(\e_5)M(\e_6)\rg$ relevant for the computation of $X_5$.
For more details about the actual calculation, I refer to Appendix B.

In Fig.\ref{Fig4}a, the results for $\chi_3^{(3)}(\om)$ and $\chi_5^{(3)}(\om)$ are shown for $n=0$ and two different temperatures. 
\begin{figure}[h!]
\centering
\includegraphics[width=7.5cm]{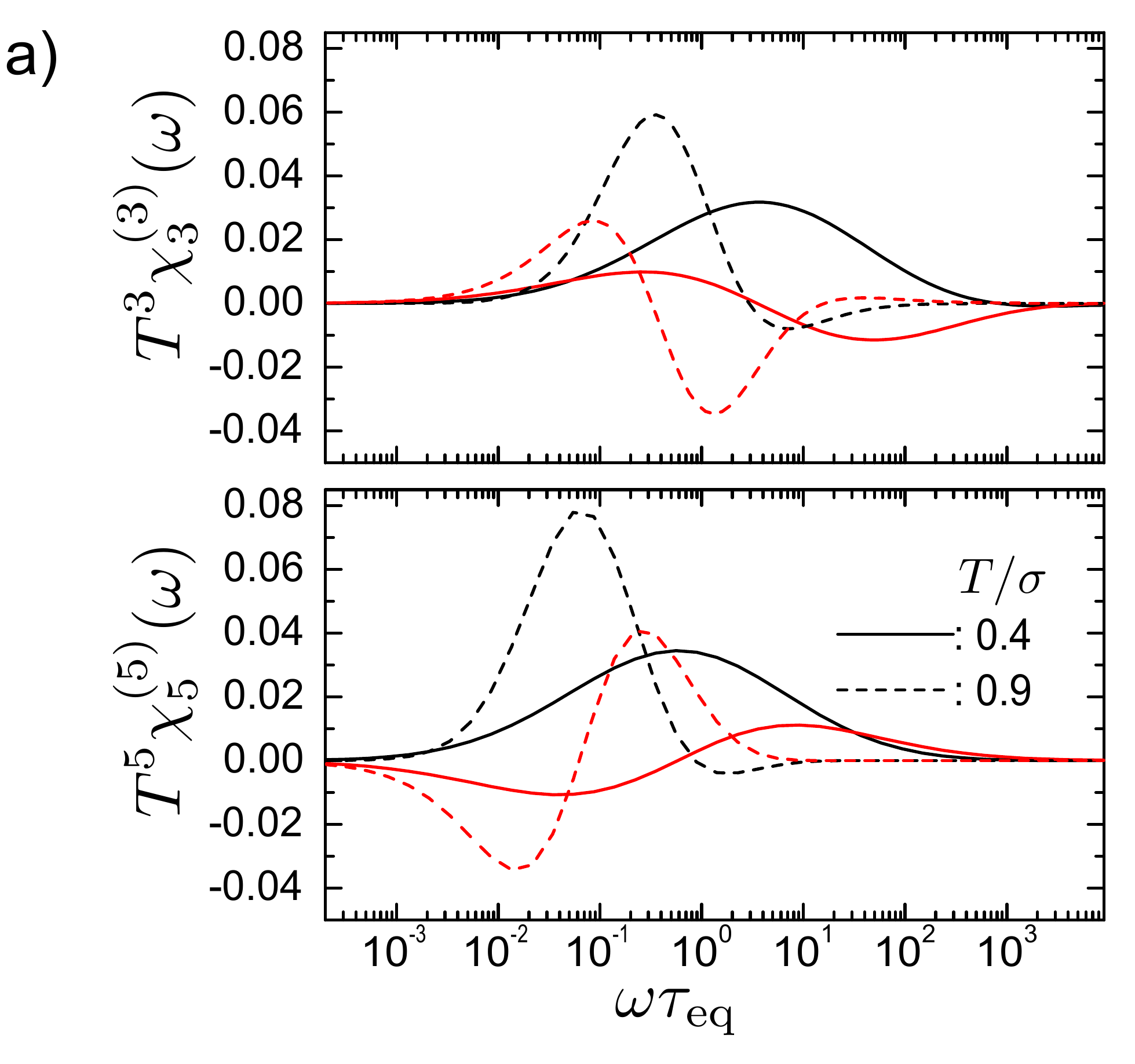}
\vspace{-0.25cm}
\hspace{0.4cm}
\includegraphics[width=8.0cm]{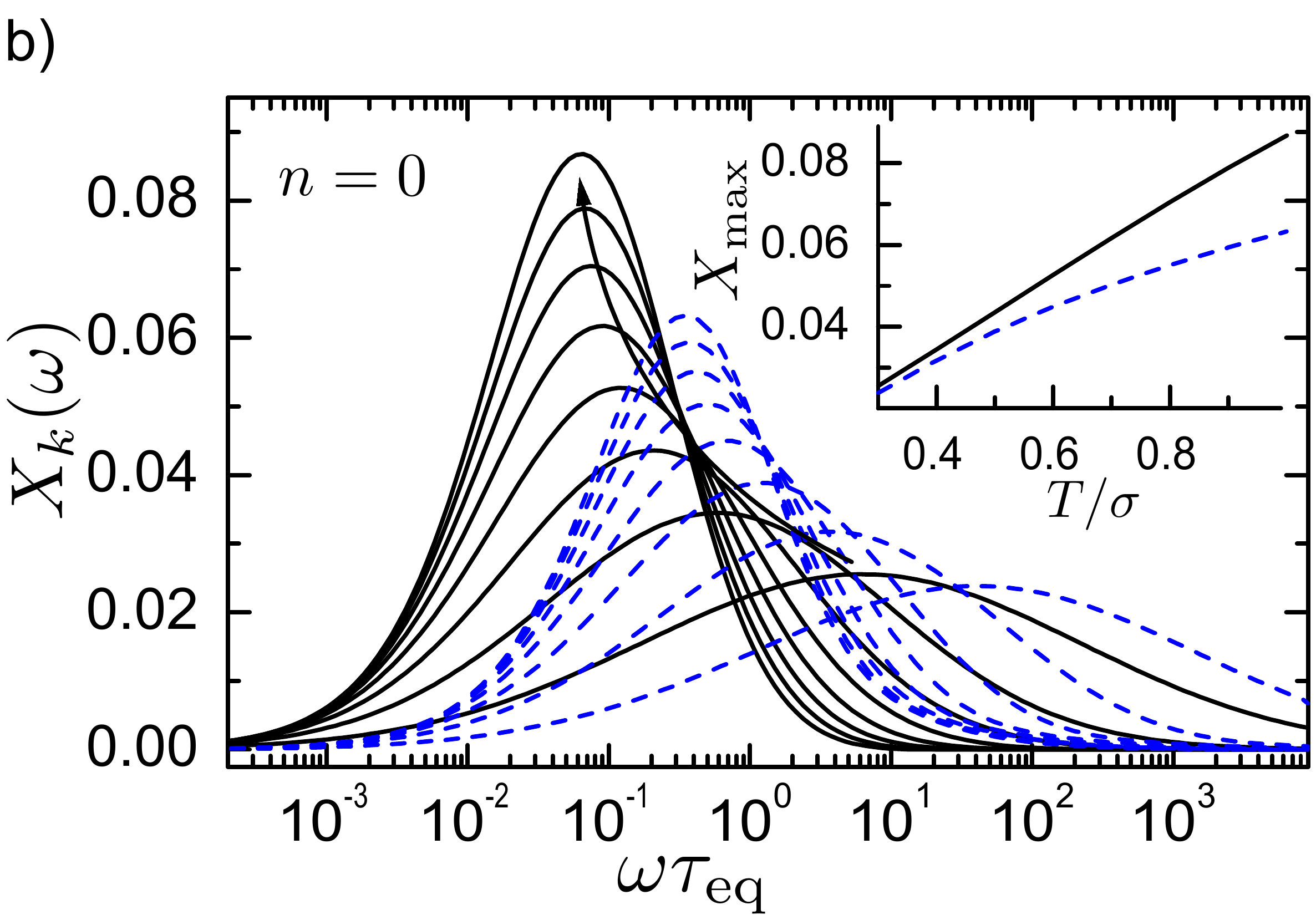}
\vspace{-0.25cm}
\caption{{\bf a)}$\chi_3^{(3)}(\om)$ (upper panel) and $\chi_5^{(5)}(\om)$ (lower panel) for $n=0$ and two different temperatures. 
The red lines represent the imaginary part and the black lines the real part.
{\bf b)} $X_k(\om)$ as a function of frequency for temperatures ranging from $T=0.3\s$ to $T=\s$ as indicated by the arrow. Black lines: $X_5(\om)$, dashed blue lines: $X_3(\om)$.
The inset shows the temperature dependence of the hump height. $\t_{\rm eq}$ denotes the mean relaxation time, given by 
$\t_{\rm eq}=\k_\infty^{-1}e^{(3/2)\b^2\s^2}$, cf. I.
}
\label{Fig4}
\end{figure}
It is evident, that the overall behavior is very similar.
Fig.\ref{Fig4}b shows the moduli $X_k(\om)$ as a function of temperature.
Both, $X_3$ and $X_5$ show a hump, the height of which increases with increasing temperature, cf. the inset of 
Fig.\ref{Fig4}b.
Thus, for $n=0$, i.e. energy independent variables, a similar behavior for the nonlinear responses is observed.
Compared to the experimental observations, in particular the temperature dependence of the height of the humps is different.

As in I, the next step consists in varying $n$, which means to consider different dynamical variables.
For $n=1$, the results for $X_3$ and $X_5$ differ because for $X_3$ it has been observed that a hump only exists above a certain threshold temperature and this is basically temperature-independent.
This does not hold for $X_5$, as shown in Fig.\ref{Fig5}.
\begin{figure}[h!]
\centering
\includegraphics[width=7.7cm]{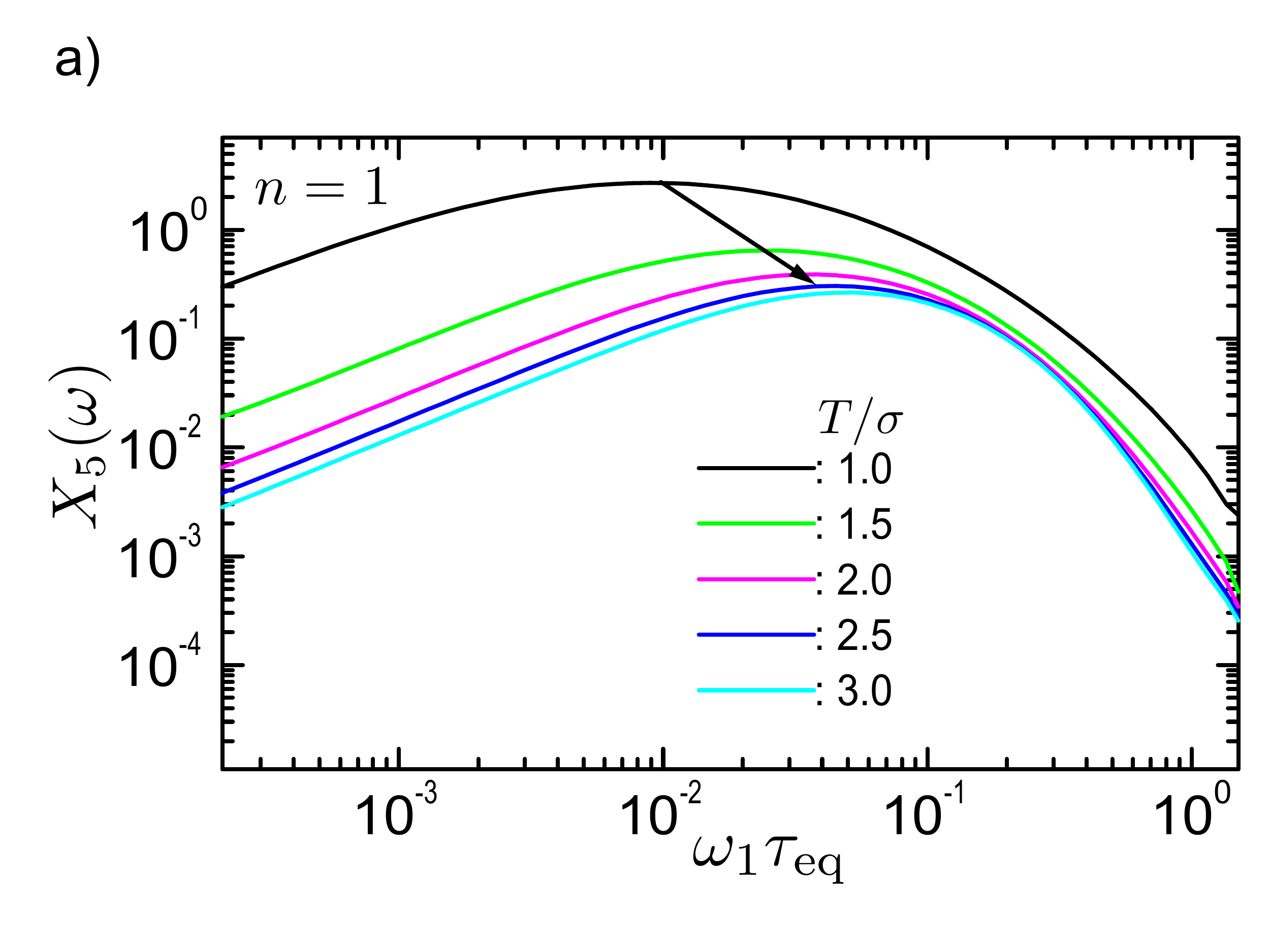}
\vspace{-0.25cm}
\hspace{0.4cm}
\includegraphics[width=7.5cm]{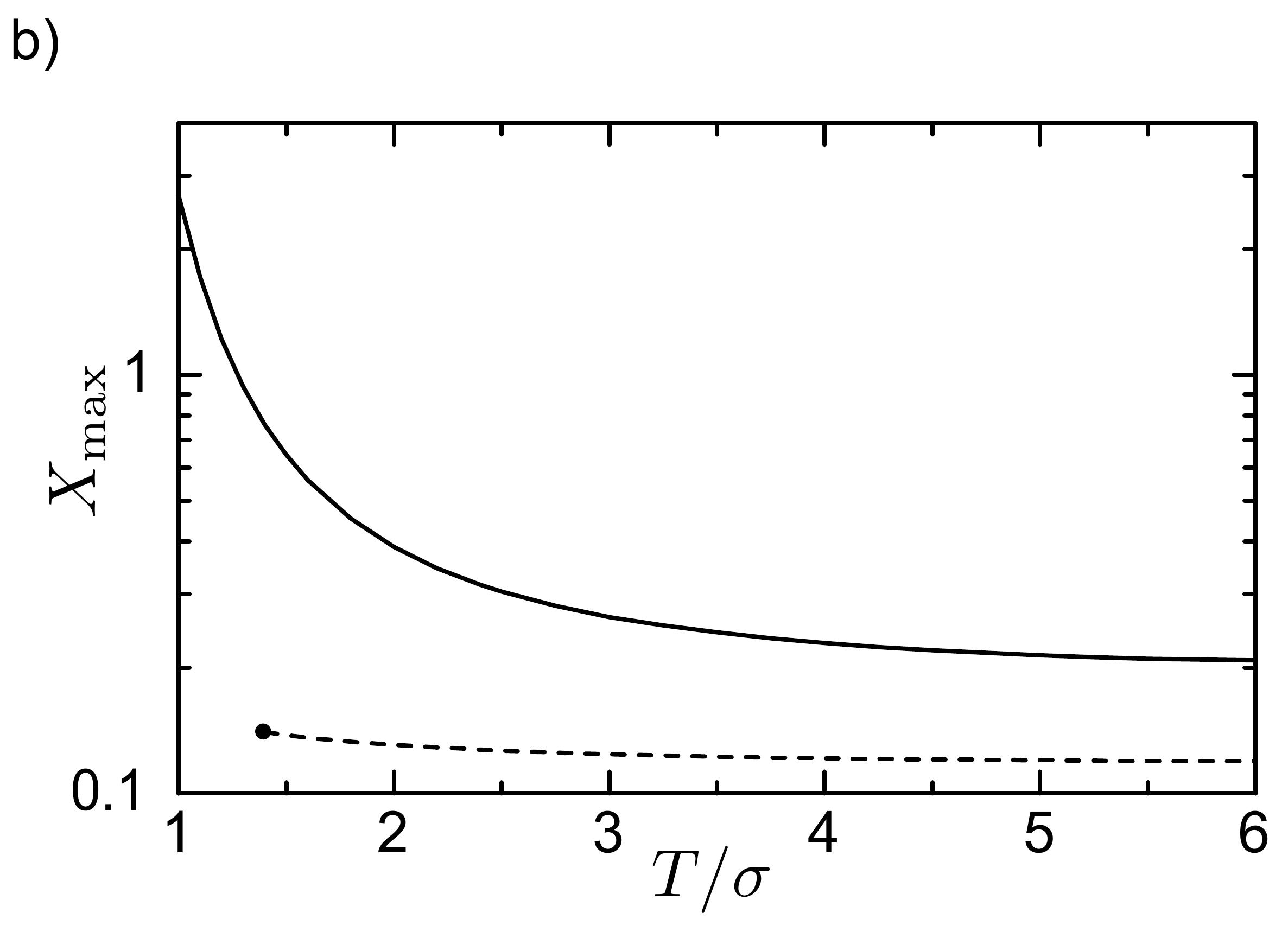}
\vspace{-0.25cm}
\caption{{\bf a)}$X_5(\om)$ for $n=1$ and different temperatures as indicated. 
Here, the scaled frequency is given by $\om_n=\om e^{n\b^2\s^2}$.
{\bf b)} Temperature dependence of the hump height. Full line: $X_5$, dashed line: $X_3$.
}
\label{Fig5}
\end{figure}
Here, the maximum height decreases as a function of temperature and it becomes very large for low temperatures.
Therefore, for this choice, $n=1$, the humps observed in the $X_k$ behave differently at low temperatures and this only changes gradually with increasing temperature.

In I, $X_3$ also for $n=-1$ was considered, in which case a hump is observed that becomes smaller for increasing temperature.
Without showing the results here, it is noticed that a similar behavior is found for $X_5$ with the relative height of the hump for $X_5$ being larger than for the third-order response.
However, it was already mentioned in I that the case $n=-1$ is special in the sense that the mean relaxation time is temperature independent.
Thus, I will not discuss this case further.

Another interesting choice is given by $n=-4$ because in this case the mean relaxation time 
$\lg\t^{(n)}\rg=\k_\infty^{-1} e^{(n+1)(n+3)/2\b^2\s^2}$ concides with $\t_{\rm eq}$, the corresponding quantity for $n=0$.
In Fig.\ref{Fig6}, the moduli $X_k(\om)$ (a) and the height of the humps (b) are plotted.
\begin{figure}[h!]
\centering
\includegraphics[width=7.5cm]{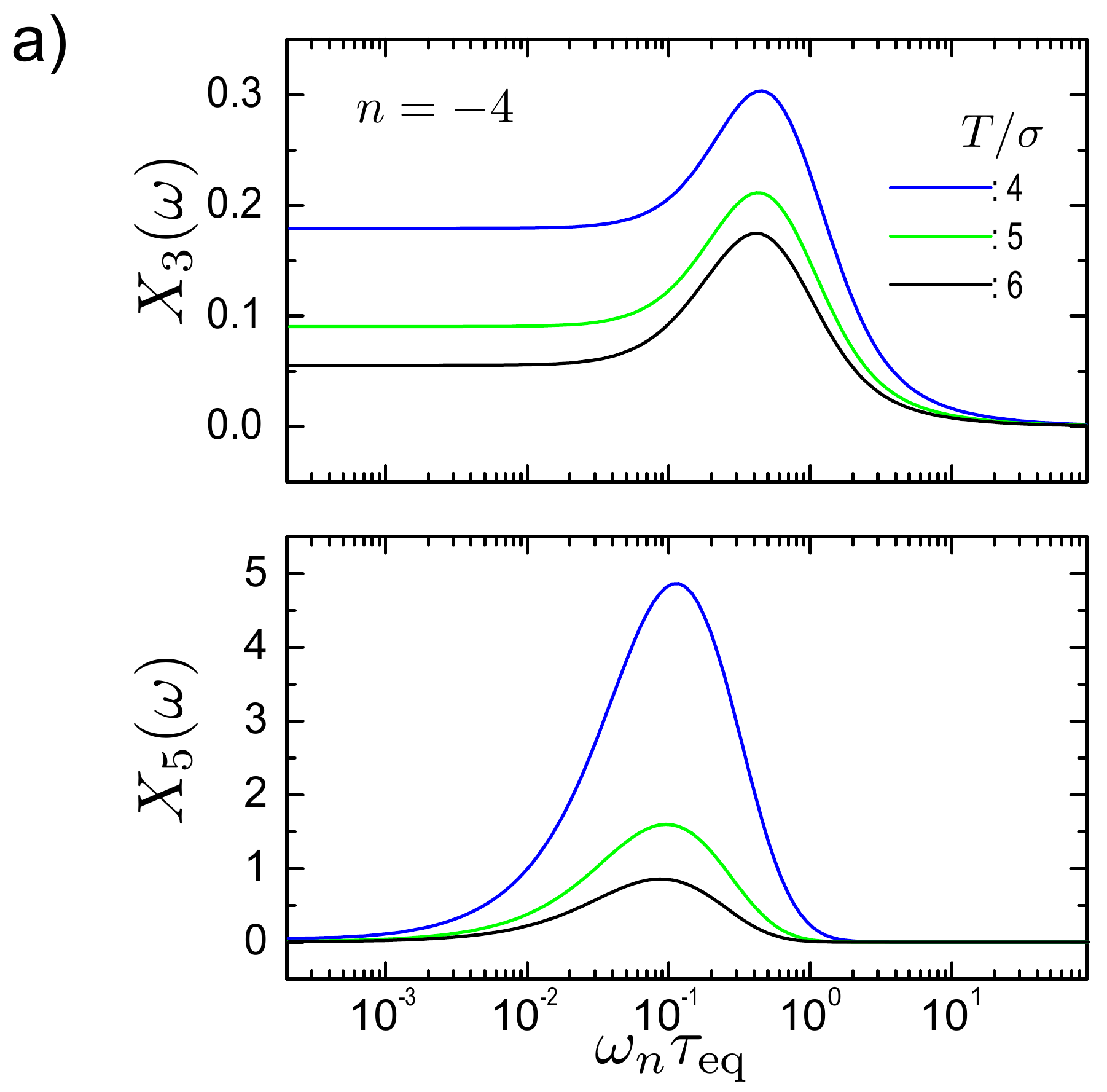}
\vspace{-0.25cm}
\hspace{0.4cm}
\includegraphics[width=8.0cm]{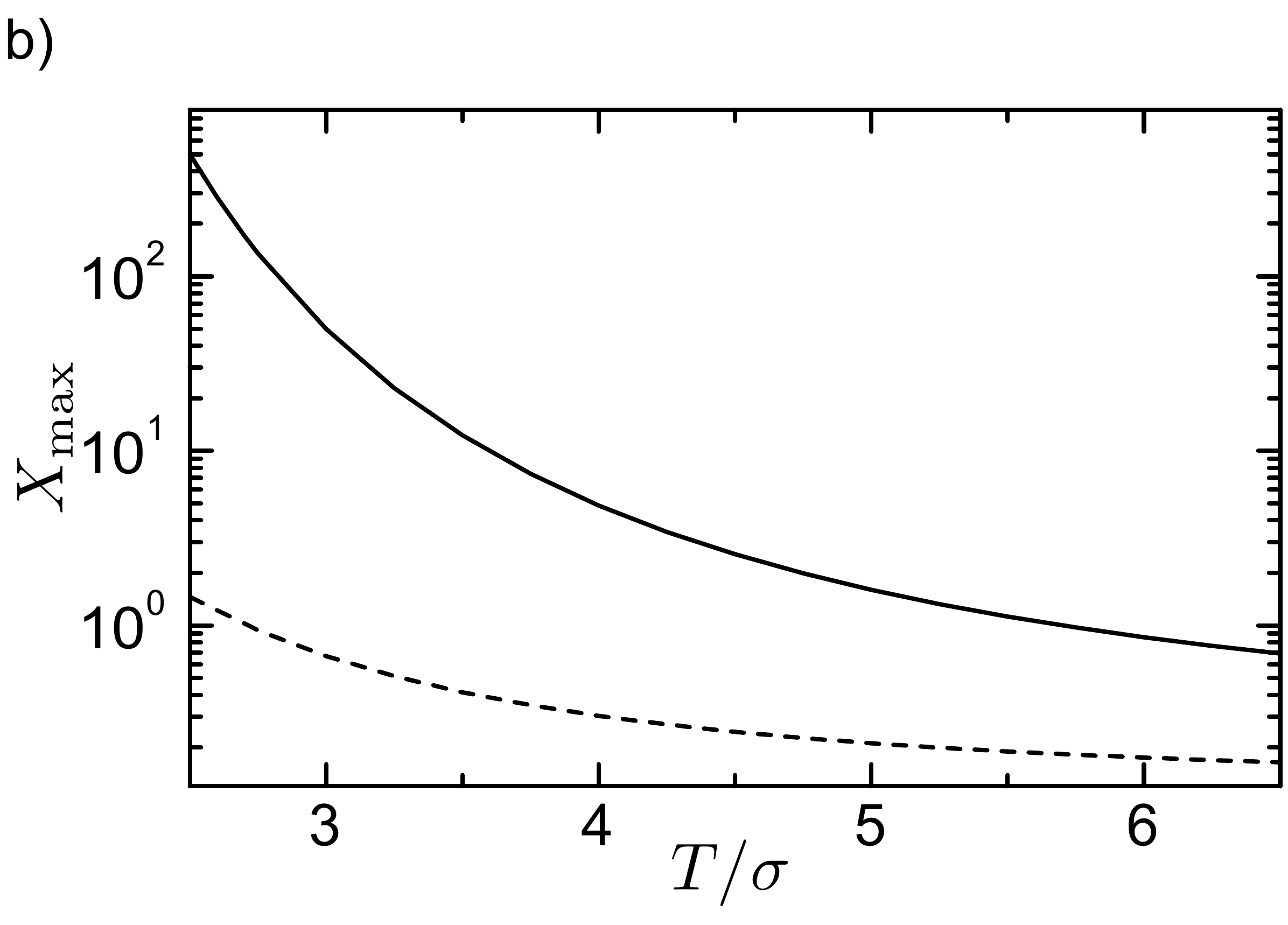}
\vspace{-0.25cm}
\caption{{\bf a)}$X_3(\om)$ (upper panel) and $X_5(\om)$ (lower panel) for $n=-4$ and different temperatures. 
{\bf b)} Temperature dependence of the hump height.
}
\label{Fig6}
\end{figure}
For this case, it is evident, that both nonlinear response functions show a behavior exhibiting a hump with a temperature dependent height that is decreasing with increasing temperature.

As discussed in detail in I for $X_3$, it is the low-frequency limit (cf. eq.(\ref{Chi55.0})) that determines the existence of a hump-like structure also for $X_5$. 
These results obtained for $X_5$ show that different scenarios are possible ranging from a similar to a very different behavior of the different $X_k$.
\section*{V. Conclusions}
In the present paper I have extended the calculation of nonlinear response functions for Markov processes presented in I\cite{G75} to the fifth-order response.
In particular, I have considered the same examples of simple stochastic models, namely the ADWP model and the trap model with a Gaussian density of states.

For the ADWP model, I find that in the temperature range where $X_3$ shows a significant hump with a height that decreases with temperature the fifth-order response $X_5$ shows either trivial or the opposite behavior.
Only for temperatures that are also higher than the higher characteristic temperature for $X_5$, one finds a decreasing height of the corresponding hump.
In that temperature regime, however, the hump in $X_3$ will be very small or vanishes.
It is therefore unlikely that the ADWP model in the present form could to be applicable in a straightforward way for the interpretation of the dielectric relaxation in the investigated systems.

In case of the trap model, a strong dependence of the behavior of $X_5$ on the choice of the dynamical variable coupling to the field is observed.
This situation is similar to the corresponding one for $X_3$.
For various values of the parameter $n$ the overall behavior of the hump is found to be similar for $X_3$ and $X_5$.
If one chooses $n=1$, however, one even finds a different temperature dependence of the corresponding heights, cf. Fig.\ref{Fig5}b.
A choice that might resemble the experimental findings partially would be $n=-4$ although no comparison is attempted in the present paper.

A note regarding the nature of the models considered appears in order.
In the present paper (and also in I) only (mean-field) models for relaxation with a well-defined stochastic dynamics have been considered.
In particular, no assumptions about the relation between the relaxation times and thermodynamic quantities (such as configurational entropy) have been made.
Furthermore, I did not not assume any distributions of relaxation times.
In phenomenological models, often a relation like the one assumed in the Adam-Gibbs model is assumed to hold\cite{Kim:2016}.
This is a perfectly valid approach to obtain a meaningful parameterization of experimental data.
However, from a theoretical point of view it appears more sound to start from given dynamical rules and calculate experimentally relevant quantities on that basis.

Calculations such as the ones performed in the preceeding sections can be further applied to other nonlinear sequences of external fields as they have been considered recently\cite{LHote:2014,YoungGonzales:2015}.
Furthermore, in the framework of models similar to the ones considered here, it is possible to compare the response of the system after a temperature-jump with the effects of strong external fields.
In a forthcoming publication, the analysis of higher-order nonlinear response functions will be performed for stochastic models obeying a Langevin equation instead of a master equation and the results will be compared.

The conclusion to be drawn from the calculations presented in the present paper is that the models considered here can only be used to describe the higher-order nonlinear response functions of supercooled liquids as observed experimentally for rather limited sets of parameters.
This means that these response functions might be of great value when it comes to discriminate among various models for the dynamics.
\section*{Acknowledgment}
I thank Roland B\"ohmer, Gerald Hinze, Francois Ladieu and Jeppe Dyre for fruitful discussions.
\begin{appendix}
\section*{Appendix A: Nonlinear response functions for the ADWP model}
\setcounter{equation}{0}
\renewcommand{\theequation}{A.\arabic{equation}}

$\chi_3^{(\a)}(\om)$:

\noindent
Here, I repeat the results for the third-order response for the two-state ADWP model given in I\cite{G75}.
One has for $\a=1$, $3$:
\be\label{Chi3.ADWP}
\chi_3^{(\a)}(\om)=\D\chi_3\times S_3^{(\a)}(\om\t)={M^4\over20}\b^3\left(1-\d^2\right)\times S_3^{(\a)}(\om\t)
\ee
with $S_3^{(\a)}(x)$ only depending on the product $x=\om\t$ and given explicitly in eq.(15) in I\cite{G75}.

When compared to the model of Brownian rotational diffusion, the following can be observed. 
For $\D=0$, $\chi_3^{(\a)}(\om)$ is very similar to the corresponding expression for the model of rotational Brownian motion.
For finite $\D$, however, the third-order response for the ADWP-model shows a characteristic temperature dependence, that is absent in the latter model.

The static nonlinear susceptibilites are determined by the limiting values of the spectral function, $S_3^{(\a)}(0)=(3\d^2-1)/\a$, and thus is given by:
\be\label{Chi3.0.ADWP}
\chi_3^{(\a)}(0)=\left({M^4\over20\a}\right)\b^3\left(3\d^2-1\right)\left(1-\d^2\right)
\ee
I note, that $\chi_3^{(\a)}(0)$ is determined by the fourth-order cumulant, given here in terms of the central moments 
$\mu_n=\lg(M-\lg M\rg)^n\rg$,
\[
\chi_3^{(\a)}(0)\sim\k_4(M)=\mu_4-3\mu_2^2
=2M^4\left(3\d^2-1\right)\left(1-\d^2\right).
\]
$\chi_{2;1}(\om)$:

\noindent
Here, for completeness and the convenience of the reader I give the spectral function for the response that is quadratic in a dc field and linear in an ac field with frequency $\om$, a situation that has been considered in detail in ref.\cite{LHote:2014}.
Again denoting $x=\om\t$, one finds in a calculation that is competely equivalent to the one performed in the case of an ac field:
\be\label{S33.ADWP}
S_{2;1}(x)=\d^2{(3-x^2)+ix(6+2x^2+x^4)\over(1+x^2)^3}-{2+ix\over2(1+x^2)^2}
\ee
This spectral function has limiting values 
\[
S_{2;1}(0)=3\d^2-2
\quad\mbox{and}\quad
S_{2;1}(\infty)=0
\]

\noindent
$\chi_5^{(5)}(\om)$:

\noindent
Here, I give the result for the spectral function determining the frequency-dependence of the $5\om$-component of the fifth-order response to an ac field. The calculation is performed as outlined in Section II, using $\mu=0$ and  $\g=1$, but the results are independent of this particular choice for a two-state model.
One finds:
\Be\label{S5.ADWP}
S_5(x)
&&\hspace{-0.6cm}=
{1\over15N(x)}
\left\{(2-15\!\cdot\!\d^2+15\!\cdot\!\d^4)-5(6-155\!\cdot\!\d^2+255\!\cdot\!\d^4)x^2\right.
\nonumber\\
&&\hspace{1.5cm}
\left.+2(-612+3445\!\cdot\!\d^2+2055\!\cdot\!\d^4)x^4\right.
\nonumber\\
&&\hspace{1.5cm}
\left.
-20(176+865\!\cdot\!\d^2)x^6+3072x^8\right\}
\\
&&\hspace{-0.3cm}
+
{ix\over8N(x)}
\left\{(11-104\!\cdot\!\d^2+120\!\cdot\!\d^4)+10(17+4\!\cdot\!\d^2-180\!\cdot\!\d^4)x^2\right.
\nonumber\\
&&\hspace{1.5cm}
\left.+(-293+9424\!\cdot\!\d^2+960\!\cdot\!\d^4)x^4\right.
\nonumber\\
&&\hspace{1.5cm}
\left.
-20(157+160\!\cdot\!\d^2)x^6+192x^8\right\}
\nonumber
\Ee
with the denominator given by
\be\label{Nx.ADWP}
N(x)=(1+x^2)(1+4x^2)(1+9x^2)(1+16x^2)(1+25x^2)
\ee
Similar to the situation for the third-order response, one has
\Be\label{Ch50.Kum6}
\chi_5^{(5)}(0)\sim\k_6(M)
&&\hspace{-0.5cm}=
\mu_6-15\mu_4\mu_2-10\mu_3^2+30\mu_2^3
\nonumber\\
&&\hspace{-0.5cm}=
8M^6\left(2-15\d^2+15\d^4\right)\left(1-\d^2\right).
\nonumber
\Ee
\section*{Appendix B: Fifth-order response functions for a trap model}
\setcounter{equation}{0}
\renewcommand{\theequation}{B.\arabic{equation}}
In the present Appendix, I will give the expression for $\chi_5^{(5)}(\om)$ for the Gaussian trap model also considered in I. 
The calculation is performed as outlined in Section II and also in App.B of I.
As the results for the third-order response only weakly depend on the values of $\mu$ and $\g$ in eq.(\ref{Wkl.HX}), I restrict the calculation of the fifth-order response to the case $\mu=1$.
This means that the coupling of the field takes place only via the initial state of the transition.
In view of the properties of the trap model this is a meaningful assumption because the field-free transition rates given by eq.(\ref{W.trap}) depend on the energy of the initial state of the transition but not on the energy of the destination state.
Additionally, I again use the Gaussian factorization approximation for the variable $M(\e)$ as has been employed in eq.(22) of I but now applied to products of the form $\lg M(\e_1)M(\e_2)M(\e_3)M(\e_4)M(\e_5)M(\e_6)\rg$, yielding 15 independent terms if one assumes 
$\lg M(\e)\rg=0$ and $\lg M(\e_1)M(\e_2)\rg=\d(\e_1-\e_2)e^{-n\b\e_1}$, cf. eq.(\ref{M.eps.gauss}).

Using a discrete notation (as in Appendix B of I), i.e. $G_{kl}(t)=G(\e_k,t|\e_l,0)$, $p^{eq}_k=p^{eq}(\e_k)$, $\rho_k=\rho(\e_k)$ and
$\k_k=\k(\e_k)$, one finds the following expression for the fifth-order response:
\Be\label{chi5.allg}
\chi_5^{(5)}(\om)=
{\lg\k\rg\over32}\b^5\left\{\chi_{k}(\om)+\chi_{kl}(\om)+\chi_{klm}(\om)\right\}
\Ee
where the individual terms are given by:
\Be\label{chi.k}
\chi_{k}(\om)
&&\hspace{-0.6cm}=
3\sum_k\rho_k\lg M_k^2\rg \left[5\lg M_k^2\rg^2S_{\a;k}(\om)+\lg M_k^2\rg\{\overline{\lg M^2\rg}-5\lg M_k^2\rg\}S_{\b;k}(\om)\right.
\\
&&\hspace{2.5cm}
+\left.
\{\overline{\lg M^2\rg^2}-5\lg M_k^2\rg^2\}S_{\g;k}(\om)
\right]
\nonumber
\Ee
\Be\label{chi.kl}
\chi_{kl}(\om)
&&\hspace{-0.6cm}=
3\sum_{k,l}\rho_k\rho_l\lg M_k^2\rg\lg M_l^2\rg \left[\lg M_k^2\rg S_{\a;kl}(\om)+\lg M_l^2\rg S_{\b;kl}(\om)\right.
\\
&&\hspace{4cm}
+\left.
{1\over3}\{\overline{\lg M^2\rg}-3\lg M_l^2\rg\}S_{\g;kl}(\om)
\right]
\nonumber
\Ee
\be\label{chi.klm}
\chi_{klm}(\om)=\sum_{k,l,m}\rho_k\rho_l\rho_m\lg M_k^2\rg\lg M_l^2\rg\lg M_m^2\rg S_{klm}(\om)
\ee
where I used the following definitions
\be\label{kap.M2hx.mit}
\lg\k\rg=\sum_kp^{eq}_k\k_k
\quad\mbox{and}\quad
\overline{\lg M^2\rg^x}=\sum_k\rho_k\lg M_k^2\rg^x
\,\,;\,\,x=1,2
\ee

The spectral functions are defined by:
\Be\label{S.a.k}
S_{\a;k}(\om)
&&\hspace{-0.6cm}=
{1\over120}S_k(\om)-{\k_k\over24}\left[S_{kk}(4\om,\om)+2S_{kk}(2\om,3\om)\right]
\nonumber\\
&&\hspace{-0.1cm}
+{\k_k^2\over4} S_{kkk}(2\om,2\om,\om)
\nonumber\\
&&\hspace{-0.1cm}
+{\k_k^2\over6}\left[S_{kkk}(3\om,\om,\om)+S_{kkk}(\om,3\om,\om)+S_{kkk}(\om,\om,3\om)\right]
\\
&&\hspace{-0.1cm}
-{\k_k^3\over2}\left[S_{kkkk}(2\om,\om,\om,\om)+S_{kkkk}(\om,2\om,\om,\om)+S_{kkkk}(\om,\om,2\om,\om)\right]
\nonumber\\
&&\hspace{-0.1cm}
+\k_k^4S_{kkkkk}(\om)
\nonumber
\Ee
\Be\label{S.b.k}
S_{\b;k}(\om)
&&\hspace{-0.6cm}=
{\k_k\over12}S_{kk}(3\om,2\om)
-{\k_k^2\over4}\left[S_{kkk}(2\om,\om,2\om)+S_{kkk}(\om,2\om,2\om)\right]
\\
&&\hspace{-0.1cm}
+{\k_k^3\over2}S_{kkkk}(\om,\om,\om,2\om)
\nonumber
\Ee
\be\label{S.c.k}
S_{\g;k}(\om)={\k_k\over24}S_{kk}(\om,4\om)
\ee
\Be\label{S.a.kl}
S_{\a;kl}(\om)
&&\hspace{-0.6cm}=
-{\k_k\k_l\over6}S_{kkl}(3\om,\om,\om)
\nonumber\\
&&\hspace{-0.1cm}
+{\k_k^2\k_l\over2}\left[S_{kkkl}(2\om,\om,\om,\om)+S_{kkkl}(\om,2\om,\om,\om)\right]
\\
&&\hspace{-0.1cm}
-\k_k^3\k_lS_{kkkkl}(\om)
\nonumber
\Ee
\Be\label{S.b.kl}
S_{\b;kl}(\om)
&&\hspace{-0.6cm}=
-{\k_k\k_l\over6}\left[S_{kkl}(\om,3\om,\om)+S_{kkl}(\om,\om,3\om)\right]
\nonumber\\
&&\hspace{-0.1cm}
+{\k_k\k_l^2\over2}\left[S_{kkll}(\om,2\om,\om,\om)+S_{kkll}(\om,\om,2\om,\om)\right]
\\
&&\hspace{-0.1cm}
-\k_k\k_l^3S_{kklll}(\om)
\nonumber
\Ee
\be\label{S.c.kl}
S_{\g;kl}(\om)
={\k_k\k_l\over4}S_{kkl}(\om,2\om,2\om)-{\k_k\k_l^2\over2}S_{kkll}(\om,\om,\om,2\om)
\nonumber
\ee
\Be\label{S.kml}
S_{kml}(\om)
&&\hspace{-0.6cm}=
-{\k_k\k_m\k_l\over2}S_{kkml}(\om,2\om,\om,\om)
\\
&&\hspace{-0.1cm}
+\k_k\k_m^2\k_lS_{kkmml}(\om)
\nonumber
\Ee
with the individual spectral densities given by:
\be\label{Sk}
S_{k}(\om)={1\over\k_k+i5\om}
\ee
\be\label{Skl}
S_{kl}(\om_1,\om_2)={1\over(\k_k+i5\om)(\k_l+i\om_2)}
\ee
\be\label{Skkl}
S_{kkl}(\om_1,\om_2,\om_3)={1\over(\k_k+i5\om)(\k_k+i\om_{23})(\k_l+i\om_3)}
\ee
\be\label{Skkml}
S_{kkml}(\om_1,\om_2,\om_3,\om_4)={1\over(\k_k+i5\om)(\k_k+i\om_{234})(\k_m+i\om_{34})(\k_l+i\om_4)}
\ee
\be\label{Skkmml}
S_{kkmml}(\om)={1\over(\k_k+i5\om)(\k_k+i4\om)(\k_m+i3\om)(\k_m+i2\om)(\k_l+i\om)}
\ee
Here, $\om_{ab...}=\om_a+\om_b+...$ and all frequencies sum up to $5\om$, i.e. $\sum_n\om_n=5\om$.
From these functions all others can be obtained in a straightforward manner by replacing the corresponding indices, e.g.
$S_{kk}(\om_1,\om_2)=S_{k(l=k)}(\om_1,\om_2)$ etc..

Using the above results, it is straightforward to compute the zero-frequency limit of the fifth-order response which is given by:
\be\label{Chi55.0}
\chi_5^{(5)}(0)={1\over128}\left( \overline{\lg M^2\rg^3_T}-2\overline{\lg M^2\rg^2_T}\cdot\overline{\lg M^2\rg}
-\overline{\lg M^2\rg_T}\cdot\overline{\lg M^2\rg^2}
+2\overline{\lg M^2\rg_T}\cdot(\overline{\lg M^2\rg})^2
\right)
\ee
with
\be\label{Mhn.inf}
\overline{\lg M^2\rg^x_T}=\sum_kp^{eq}_k\lg M_k^2\rg^x
\ee
the high-temperature limit of which coincides with $\overline{\lg M^2\rg^x}$ according to eq.(\ref{kap.M2hx.mit}) because of 
$p_k^{eq}\to\rho_k$ for $T\to\infty$.
\end{appendix}

\begin{thebibliography}{10}

\bibitem{CrausteThibierge10}
C.~Crauste-Thibierge et~al.,
\newblock Phys. Rev. Lett. {\bf 104}, 165703 (2010).

\bibitem{Brun11}
C.~Brun et~al.,
\newblock Phys. Rev. B {\bf 84}, 104204 (2011).

\bibitem{Lunki17}
P.~Lunkenheimer, M.~Michl, T.~Bauer, and A.~Loidl,
\newblock arXiv , 1704.07348 (2017).

\bibitem{Bouchaud05}
J.~Bouchaud and G.~Biroli,
\newblock Phys. Rev. B {\bf 72}, 064204 (2005).

\bibitem{SRS91}
K.~Schmidt-Rohr and H.~Spiess,
\newblock Phys. Rev. Lett. {\bf 66}, 3020 (1991).

\bibitem{HWZS95}
A.~Heuer, M.~Wilhelm, H.~Zimmermann, and H.~Spiess,
\newblock Phys. Rev. Lett. {\bf 75}, 2851 (1995).

\bibitem{G23}
R.~B\"ohmer et~al.,
\newblock J. Non-Cryst. Solids {\bf 235-237}, 1 (1998).

\bibitem{Sillescu99}
H.~Sillescu,
\newblock J.Non-Cryst. Solids {\bf 243}, 81 (1999).

\bibitem{Ediger00}
M.~Ediger,
\newblock Annu. Rev. Phys. Chem. {\bf 51}, 99 (2000).

\bibitem{Richert02}
R.~Richert,
\newblock J. Phys.: Condens. Matter {\bf 14}, R703 (2002).

\bibitem{Toninelli05}
C.~Toninelli, M.~Wyart, L.~Berthier, G.~Biroli, and J.~Bouchaud,
\newblock Phys. Rev. E {\bf 71}, 041505 (2005).

\bibitem{Berthier:2011}
L.~Berthier,
\newblock Physics {\bf 4}, 42 (2011).

\bibitem{Bauer:2013}
T.~Bauer, P.~Lunkenheimer, and A.~Loidl,
\newblock Phys. Rev. Lett. {\bf 111}, 225702 (2013).

\bibitem{Casalini:2015}
R.~Casalini, D.~Fragiadakis, and C.~M. Roland,
\newblock J. Chem. Phys. {\bf 142}, 064504 (2015).

\bibitem{Michl:2015}
M.~Michl, T.~Bauer, P.~Lunkenheimer, and A.~Loidl,
\newblock Phys. Rev. Lett. {\bf 114}, 067601 (2015).

\bibitem{Michl:2016}
M.~Michl, T.~Bauer, P.~Lunkenheimer, and A.~Loidl,
\newblock J. Chem. Phys. {\bf 144}, 114506 (2016).

\bibitem{Patro:2017}
L.~N. Patro, O.~Burghaus, and B.~Roling,
\newblock J. Chem. Phys. {\bf 146}, 154503 (2017).

\bibitem{Brun11b}
C.~Brun, C.~Crauste-Thibierge, F.~Ladieu, and D.~L'Hote,
\newblock J. Chem. Phys. {\bf 134}, 194507 (2011).

\bibitem{Pick:2015}
R.~M. Pick,
\newblock J. Chem. Phys. {\bf 142}, 064511 (2015).

\bibitem{Richert:2016}
R.~Richert,
\newblock J. Chem. Phys. {\bf 144}, 114501 (2016).

\bibitem{Ladieu:2012}
F.~Ladieu, C.~Brun, and D.~L'h{\^o}te,
\newblock Phys. Rev. B {\bf 85}, 184207 (2012).

\bibitem{Morita86}
A.~Morita,
\newblock Phys. Rev. A {\bf 34}, 1499 (1986).

\bibitem{DD95}
J.~Dejardin and G.~Debiais,
\newblock Adv. Chem. Phys. {\bf 91}, 241 (1995).

\bibitem{Dejardin00}
J.~Dejardin and Y.~Kalmykov,
\newblock Phys. Rev. E {\bf 61}, 1211 (2000).

\bibitem{Kalmykov01}
Y.~Kalmykov,
\newblock Phys. Rev. E {\bf 65}, 021101 (2001).

\bibitem{Dejardin:2014}
P.~M. D{\'e}jardin and F.~Ladieu,
\newblock J. Chem. Phys. {\bf 140}, 034506 (2014).

\bibitem{G75}
G.~Diezemann,
\newblock Phys. Rev. E {\bf 85}, 051502 (2012).

\bibitem{G39}
G.~Diezemann,
\newblock Europhys. Lett. {\bf 53}, 604 (2001).

\bibitem{G46}
R.~B\"ohmer and G.~Diezemann,
\newblock {\em in: Broadband Dielectric Spectroscopy},
\newblock Springer, Berlin, Heidelberg, New York, 2002.

\bibitem{Dyre95}
J.~Dyre,
\newblock Phys. Rev. B {\bf 51}, 12276 (1995).

\bibitem{MB96}
C.~Monthus and J.~Bouchaud,
\newblock J. Phys. A-Math. Gen. {\bf 29}, 3847 (1996).

\bibitem{G64}
G.~Diezemann,
\newblock J. Phys.: Condens. Mat. {\bf 19}, 205107 (2007).

\bibitem{G71}
C.~Rehwald et~al.,
\newblock Phys. Rev. E {\bf 82}, 021503 (2010).

\bibitem{G73}
G.~Diezemann and A.~Heuer,
\newblock Phys. Rev. E {\bf 83}, 031505 (2011).

\bibitem{Albert:2016}
S.~Albert et~al.,
\newblock Science {\bf 352}, 1308 (2016).

\bibitem{SBLC96}
B.~Schiener, R.~B\"ohmer, A.~Loidl, and R.~Chamberlin,
\newblock Science {\bf 274}, 752 (1996).

\bibitem{G16}
B.~Schiener, R.~Chamberlin, G.~Diezemann, and R.~B\"ohmer,
\newblock J. Chem. Phys. {\bf 107}, 7746 (1997).

\bibitem{vkamp81}
N.~van Kampen,
\newblock {\em Stochastic Processes in Physics and Chemistry},
\newblock North-Holland, Amsterdam, 1981.

\bibitem{CR03}
A.~Crisanti and F.~Ritort,
\newblock J. Phys. A-Math. Gen. {\bf 36}, R181 (2003).

\bibitem{G54}
G.~Diezemann,
\newblock Phys. Rev. E {\bf 72}, 011104 (2005).

\bibitem{LHote:2014}
D.~L'Hote, R.~Tourbot, F.~Ladieu, and P.~Gadige,
\newblock Phys. Rev. B {\bf 90}, 104202 (2014).

\bibitem{FS02}
S.~Fielding and P.~Sollich,
\newblock Phys. Rev. Lett. {\bf 88}, 050603 (2002).

\bibitem{Kim:2016}
P.~Kim, A.~R. Young-Gonzales, and R.~Richert,
\newblock J. Chem. Phys. {\bf 145}, 064510 (2016).

\bibitem{YoungGonzales:2015}
A.~R. Young-Gonzales, S.~Samanta, and R.~Richert,
\newblock J. Chem. Phys. {\bf 143}, 104504 (2015).

\end{thebibliography}
\end{document}